\documentclass[%
 reprint,
 prc,
 amsmath,amssymb,
 aps,
 lengthcheck,
floatfix,
]{revtex4}

\usepackage{graphicx}
\usepackage{dcolumn}
\usepackage{bm}
\usepackage{hyperref}
\usepackage[utf8]{inputenc}
\usepackage{multirow}
\usepackage{soul}
\usepackage{float}
\usepackage{graphicx}
\usepackage{array}
\usepackage{tabularx}
\usepackage{times}
\usepackage[normalem]{ulem}
\usepackage{color}
\usepackage{adjustbox}
\usepackage{cellspace}

\usepackage{lipsum}

\newcommand{\be}{\begin{equation}}
\newcommand{\ee}{\end{equation}}
\newcommand{\bea}{\begin{eqnarray}}
\newcommand{\eea}{\end{eqnarray}}

\newcommand{\comment}[1]{}
\renewcommand\sout{\bgroup \color{red} \ULdepth=-.5ex \ULset}
\def\simge{\mathrel{\rlap{\raise 0.511ex
     \hbox{$>$}}{\lower 0.511ex \hbox{$\sim$}}}}
\def\simle{\mathrel{\rlap{\raise 0.511ex
      \hbox{$<$}}{\lower 0.511ex \hbox{$\sim$}}}}

\begin{document}


\title{Implications of $\sigma$-cut potential on Antikaon condensates in neutron stars}

\author{Prashant Thakur$^1$}
\author{Yashmitha Kumaran$^2$}
\author{Lakshana Sudarsan$^{3,5}$}
\author{Krishna Kunnampully$^{4,5}$}
\author{B. K. Sharma$^5$}
\author{T. K. Jha$^1$}
  
\affiliation{$^1$Department of Physics, BITS-Pilani, K. K. Birla Goa
Campus, Goa 403726, India\\
$^2$Instituto de Astronom\'ia, Universidad Nacional Aut\'onoma de M\'exico, 
AP 70-264, Ciudad de M\'exico 04510, M\'exico\\
$^3$Department of Mathematics, Aston School of Engineering, Aston University, Birmingham, B4 7ET, UK\\
$^4$School of Geography, Geology and the Environment College of Science and Engineering,
University of Leicester, Leicester, LE1 7RH, UK\\
$^5$Department of Physics, Amrita School of Physical Sciences, 
Amrita Vishwa Vidyapeetham, Coimbatore 641112, India}

\begin{abstract} 
We investigate the properties of neutron stars with antikaon condensation in the framework of 
the Relativistic Mean-Field (RMF) model with a $\sigma$-cut potential.
The well-known RMF models, TM1 and TM1e, are used to analyze the structure and composition of neutron stars. 
The antikaon condensation part of the equation of state (EoS) is constrained 
from the experimental data of K$^{-}$ atomic and kaon-nucleon scattering. The $\sigma$-cut potential, 
which is known to make the EoS stiffer at high densities, is modulated by a free parameter $f_{s}$. 
Our present analysis suggests that one can obtain neutron star configurations heavier than 2$M_{\odot}$ with antikaon 
condensates in most cases for $f_{s}$ = 0.6. The antikaon phase transition is a second- order for $f_{s}$ = 0.6 for
both TM1 and TM1e parameter sets. The calculated global properties of neutron stars with antikaon 
condensates i.e., mass and radius seem to be in resonable agreement with other theoretical and observational data. 
\end{abstract}


\maketitle
\section{Introduction}
Neutron stars serve as a pristine astrophysical laboratory, offering a unique opportunity to investigate nuclear matter under 
extreme densities inaccessible in terrestrial labs \cite{LAT.04,LAT.07,OER.17}. Their composition and properties are governed 
by the equation of state (EoS), which describes the behavior of matter under a wide range of densities, from a few to several 
times the nuclear saturation density ($n_{0}$). Typically, neutron stars consist mainly of neutrons with a small proportion of 
protons and electrons in $\beta$-equilibrium, ensuring charge neutrality. As neutron density rises, so does the electron density 
and momentum. At a critical density where the electron Fermi momentum matches the rest mass of the muon, the latter one starts 
appearing and may be present in an appreciable amount. Further increase in density can lead to the emergence of novel phases, 
such as hyperons \cite{AMB.60,GLE.91,WU.11,SCH.96,JHA.08,PAT.22}, kaon condensates \cite{THA.20,SHA.10}, and even 
quark matter \cite{LI.10,PAL.99}, 
in the dense core of neutron stars.

We embark on investigating the presence of antikaon condensates in dense nuclear matter and their effect on the underlying EoS 
and neutron star properties. Charge-neutral matter, primarily composed of neutrons, protons, and electrons, may also have the 
possibility of undergoing a transition to condensates with an increase in baryon density. 
The attaractive K$^{-}$-nucleon interaction
becomes stronger as the density increases, leading to a decrease in the effective mass $m_{K^{*}}$ of antikaons.
This results in a decrease in the in-medium energy of K$^{-}$-mesons, $\omega_{K^{-}}$, causing s-wave K$^{-}$ 
condensation to occur $\omega_{K^{-}}$ is equal to the K$^{-}$ chemical potential $\mu_{K^{-}}$
which is also equal to the electron chemical potential $\mu_{e}$ in a cold catalyzed (neutrino-free) neutron star matter
\cite{GLE.98}. 
Above this threshold, kaons can form a sizable population, potentially suppressing the electron population. 
Notably, the critical density for antikaon appearance is sensitive to the optical 
potential in symmetric nuclear matter. Antikaon condensation has been observed in dense baryonic matter from 
heavy-ion collisions \cite{KAP.86,NEL.87}, with subsequent studies exploring its  
occurrence in neutron stars using chiral \cite{BRO.92,LEE.94,THO.94,ELL.95} 
and relativistic mean field (RMF) models \cite{KNO.95,PRA.97}.

Glendenning and Schaffner provided a comprehensive analysis of first-order kaon condensation within the RMF model, excluding 
hyperons \cite{GLE.98}. Other investigations, employing density-dependent RMF \cite{BAN.02,GUO.03} and quark-meson coupling 
models \cite{MEN.05,RYU.07,YUE.08}, also explored kaon condensation's effects. These studies collectively suggest that kaon 
condensation is known to soften the EoS at high densities, reducing neutron star maximum mass. In contrast, recent mass 
measurements of pulsars, such as PSR J1614-2230 \cite{DEM.10,ARZ.18}, PSR J0348+0432 \cite{ANT.13}, and PSR J0740+6620 
\cite{FON.21,RIL.21}, indicate masses exceeding twice that of the Sun (M$\geq$2M$_{\odot}$). Observational data from sources 
like NICER and the GW170817 event \cite{JIA.20} constrain the EoS and neutron star properties, driving the need for models 
that incorporate these constraints effectively.

The nuclear symmetry energy $E_{sym}$ and its density dependence \cite{DUC.10,LAT.16} are crucial quantities that provide 
important information about neutron-rich matter at high density. They significantly impact the pressure of neutron
star matter and influence properties such as the radius of neutron stars \cite{CEN.09,FAT.18}. Despite their significance in 
the equation of state (EoS), the understanding of nuclear symmetry energy and its behavior at high-density
remains limited. Various studies have attempted to constrain the value of $E_{sym}$ and its high-density behavior
using data from both terrestrial experiments and astrophysical observations \cite{TSA.12,LIM.13,LAT.14,ROC.15,TEW.17}.
A recent study reported an improved neutron skin thickness of $^{208}$Pb around $(0.283 \pm 0.071)$ fm
from the Lead Radius EXperiment-II (PREX-2) \cite{ADH.21}. This measurement yielded values
of $E_{sym}$ and its slope $L$ at nuclear saturation density $n_{0}$ of $(38.1 \pm 4.7)$ MeV and
$(106 \pm 37)$, respectively \cite{REE.21}, which are higher than previously reported values \cite{OER.17}.
These updated values were derived from comparisons of experimental data from finite nuclei and heavy-
ion collisions with various microscopic model calculations. The models that we use and compare in the
present work, TM1e \cite{SHE.20} is an extension of the original TM1 \cite{TOKI.94}, where the $\omega$-$\rho$ cross-coupling
was introduced in the former to bring down the symmetry energy slope parameter from $L$ = 110.8 MeV to $L$ = 40 MeV, with an 
additional parameter $\Lambda_v$ = 0.0429, which agrees with the earlier reported values \cite{OER.17}.
However, TM1 has the slope parameter $L$ = 110.8 MeV which is more consistent with the recent data \cite{ADH.21}.
Therefore, these two values of $L$ cover a wide enough range to survey the effect of symmetry energy.
 
In our previous work with RMF models \cite{THAK.24}, we studied pure nucleonic and hyperon-rich neutron star matter. 
Our findings indicate that the effect of $\Lambda_v$ coupling has a more significant impact on tidal deformability compared to
the mass and radius of a neutron star when the value of $f_{s}$ is fixed. Therefore, in this work, we aim to investigate
how the $\sigma$-cut potential affects neutron star properties through $f_{s}$ and the symmetry energy through $\Lambda_v$
in the presence of antikaon condensation. 
The TM1 \cite{TOKI.94} and TM1e \cite{SHE.20} parameter sets were chosen for
the present analysis, with the TM1e set being particularly relevant due to the $\Lambda_v$ coupling which is crucial for  
tidal deformability. In ref. \cite{ZHA.18}, it is demonstrated that $f_{s}$ must be higher than 0.55 in the TM1 model 
to maintain the properties of finite nuclei without being affected by the  $\sigma$-cut potential. 
Therefore, we have set $f_{s}$= 0.6 for the present analysis. 

The criteria of large maximum neutron star masses,
small stellar radii, and a low tidal deformability as determined from the GW170817 binary NS
merger pose a significant challenge for nuclear models of the Equation of State (EoS), particularly those incorporating
non-nucleonic matter inside the neutron star. As already mentioned, the $f_{s}$ and $\Lambda_v$ coupling control
the mass, radius, and tidal deformability, respectively. The $\Lambda_v$ coupling is fixed by experimental and
observational constraints \cite{SUM.19}. The $f_{s}$= 0.6 for TM1 is fixed without changing the ground-state properties of 
finite nuclei with the original TM1 interaction and the constraints of observed massive neutron stars \cite{ZHA.18}.
In ref. \cite{ZHA.18}, they studied neutron stars with or without hyperon core, so this work focuses on
antikaon condensation in pure nucleonic neutron star matter. It is worth investigating whether the $f_{s}$= 0.6
has a similar impact on antikaon condensation as hyperonic neutron star matter.    

   
The paper is structured as follows: Section II outlines the RMF model with a $\sigma$-cut potential and the stellar 
equations for neutron stars. In Section III, we analyze the influence of the $\sigma$-cut potential on nuclear matter 
and neutron star properties with antikaon condensation. Finally, Section IV offers a summary and conclusions of our findings.

\section{Formalism}
In this section, we provide a brief description of the main features of the baryonic model used to evaluate 
the equation of state of neutron stars and the procedure for obtaining stellar properties. Specifically, 
we investigate the aspects of tidal deformability with condensates and compare them with the GW analysis.
   
\subsection{Equation of State}
We use the RMF models, namely the TM1 and TM1e to describe the equation of state (EoS) for nucleonic matter with antikaon 
condensation. The TM1e model is an extension of the TM1 parameterization with the $\Lambda_{v}$ coupling. The $\Lambda_{v}$ 
coupling term is essential to modify the density dependence of the symmetry energy \cite{BAO.14}. We also include 
$U_{cut}(\sigma)$ to make EoS stiffer at high density. The Lagrangian density \cite{SUM.19} with $U_{cut}(\sigma)$ for 
nucleonic degrees of freedom is given by:   
\begin{widetext}
\begin{eqnarray}
{\cal L}_{N}&=&\sum_{i=p,n}\overline{\Psi}_{i}\left[i\gamma_\mu
\partial^{\mu} - \left(M_{N} + g_{\sigma N}{\sigma}\right) 
-\gamma_\mu\left(g_{\omega N}\omega^{\mu}
+ \frac{g_{\rho N}}{2}\tau_{a}\rho^{a\mu}\right)
\right]{\Psi}_{i}
+\frac{1}{2}{\partial_{\mu}}{\sigma}{\partial^{\mu}}{\sigma}
-\frac{1}{2}m_{\sigma}^2{\sigma^2}-\frac{1}{3}g_{2}{\sigma^3}
-\frac{1}{4}g_{3}{\sigma^4}
\nonumber \\
&&-\frac{1}{4} {W_{\mu\nu}}{W^{\mu\nu}}
+\frac{1}{2}m_{\omega}^2{\omega_{\mu}\omega^{\mu}}+\frac{1}{4}c_{3}\left({\omega_{\mu}\omega^{\mu}}\right)^{2}
-\frac{1}{4}{R^a_{\mu\nu}}{R^{a\mu\nu}} 
+\frac{1}{2}m_{\rho}^2\rho^{a}_{\mu}\rho^{a\mu}
\nonumber \\
&&+ \Lambda_{v}\left(g^2_{\omega N}\omega_{\mu}\omega^{\mu}\right)\left(g^2_{\rho N}\rho^{a}_{\mu}\rho^{a\mu}\right)
-U_{cut}(\sigma)
\label{lag}
\end{eqnarray} 
\end{widetext}
The antisymmetric field tensors $W^{\mu\nu}$ and $R^{a\mu\nu}$ correspond to the fields $\omega^{\mu}$
and $\rho^{a\mu}$, respectively. We incorporate the $\omega$-$\rho$ coupling term as 
outlined in \cite{TOD.05}, which is crucial for altering the symmetry energy slope. In the RMF model, the meson fields are 
considered as classical fields, and the field operators are substituted with their expectation values, i.e., $\sigma$ = $\langle\sigma\rangle$, $\omega$ = $\langle\omega^{0}\rangle$ and 
$\rho$ = $\langle\rho^{30}\rangle$, 
respectively. The $U_{cut}(\sigma)$ has a logarithmic form as \cite{MAS.15}, 
which only affects the $\sigma$-field at high density \cite{ZHA.18} and is given by: 
 
\begin{eqnarray}
U_{cut}(\sigma) = \alpha \ln [ 1 + \exp\{\beta(-g_{\sigma N}\sigma/M_{N}-f_{s})\}]
\end{eqnarray}
where $\alpha$ = $m_{\pi}^{4}$ and $\beta$ = 120 \cite{MAS.15} to make the equation of state (EoS) 
stiffer at high density. The factor $f_{s}$ is a free parameter and we take $f_{s}$ = 0.6 for our calculation. As the density increases, kaon condensation occurs in the interior of neutron stars.
The Lagrangian for same is as follows:
\begin{eqnarray}
{\cal L}_{K}&=& D^{*}_{\mu}K^{*}D^{\mu}K - m^{*2}_{K}K^{*}K
\end{eqnarray}
The covariant derivative is given by $D^{\mu} = \partial_{\mu} + ig_{\omega K}\omega_{\mu} + 
i\frac{g_{\rho K}}{2}\tau_{K}.\rho_{\mu}$
and the effective mass of the kaon is $m^{*}_{K} = m_{K} + g_{\sigma K}\sigma$. 
The vector couplings $g_{\omega K}$ and $g_{\rho K}$, 
which represents the interactions between vector meson and the kaon, are determined by the SU(3) symmetry as 
$g_{\omega K}$ = $g_{\omega N}/{3}$ and $g_{\rho K}$ = $g_{\rho N}$. The scalar coupling $g_{\sigma K}$ is determined 
by the optical potenial of the $K^{-}$ in saturated nuclear matter:
\begin{eqnarray} 
U_{K^{-}}(n_{0}) = g_{\sigma K}\sigma(n_{0}) - g_{\omega K}\omega(n_{0}) 
\end{eqnarray} 
where $n_{0}$ is the symmetric nuclear matter saturation density and eq.(8) defines the kaon-nucleon interaction. It is to be 
noted that the difference between TM1 and TM1e lies in the value of $g_{\rho N}$ and $\Lambda_{v}$ couplings
as shown in Table I \cite{SHE.20}. As these couplings do not have any role to play in symmetric nuclear matter 
the kaon-nucleon interaction remains the same for both TM1 and TM1e. The coupling constants for antikaon
to the $\sigma$-meson, i.e., $g_{\sigma K}$ for various values of antikaon optical potential depths
$U_{K^{-}}$ at saturation density $n_{0}$, which is 0.145$fm^{-3}$ for TM1 are given in Table II.

Several experimental studies have demonstrated that kaons experience a repulsive interaction at saturation density 
in nuclear matter, while antikaons experience an attractive potential \cite{PAL.00,LI.97}.     
The depth of the attractive potential for antikaons is predicted to be $U_{K^{-}}(n_{0})$ = -120 MeV at $n_0$ by 
Waas and Weise \cite{WAA.97}, and $U_{K^{-}}(n_{0})$ = -100 MeV was calculated by coupled channel calculations at 
finite density \cite{KOC.94}. 
Self-consistent calculations using a chiral Lagrangian \cite{LUT.98,RAM.00} and 
coupled channel calculations, including a modified self-energy of the kaon \cite{TOL.02},
predict that the depth of the attractive potential for antikaons is approximately 
-80 MeV to -50 MeV at nuclear saturation density. According to a hybrid model \cite{FRI.99}, 
the value of the $K^{-}$ optical potential is estimated to be around 180$\pm$20 MeV.
In this study, we performed calculations using optical potentials ranging from -160 MeV to -100 MeV.    

\begin{table}
\caption{The coupling constants of the TM1 and TM1e Models.} 
{
\begin{tabular}{lllllllll} \\
\hline \hline
\multicolumn{1}{c}{Model} &
\multicolumn{1}{c}{$g_{\sigma N}$} &
\multicolumn{1}{c}{$g_{\omega N}$} &
\multicolumn{1}{c}{$g_{\rho N}$} &
\multicolumn{1}{c}{$g_{2}$$(fm^{-1})$} &
\multicolumn{1}{c}{$g_{3}$} &
\multicolumn{1}{c}{$c_{3}$} &
\multicolumn{1}{c}{$\Lambda_{v}$} &
\multicolumn{1}{c}{~}\\ \hline
TM1 & 10.0289 & 12.6139 & 9.2644 & -7.2325 & 0.6183 & 71.3075 & 0.0000 \\
TM1e& 10.0289 & 12.6139 & 13.9714 & -7.2325 & 0.6183 & 71.3075 & 0.0429 \\
\hline\hline
\end{tabular}
}
\end{table}

\begin{table}
\caption{The coupling constants for the antikaons to $\sigma$-meson i.e., $g_{\sigma K}$, for
different values of $U_{K^{-}}$ at saturation density $n_{0}$ for TM1.} 
{
\begin{tabular}{lllllllll} \\
\hline \hline
\multicolumn{1}{c}{$U_{K^{-}} (MeV)$} &
\multicolumn{1}{c}{$-100$} &
\multicolumn{1}{c}{$-120$} &
\multicolumn{1}{c}{$-140$} &
\multicolumn{1}{c}{$-160$} &
\multicolumn{1}{c}{~}\\ \hline
TM1 & 0.2537 & 0.8384 & 1.4241 & 2.0098 \\
\hline\hline
\end{tabular}
}
\end{table}
From Eq.(3), we can derive the Euler-Lagrangian equation for the kaon, as well as the 
dispersion relation for the Bose-Einstein condensation of the $K^{-}$ which is given by:
\begin{eqnarray}
\omega_{K^{-}} = m^{*}_{K} - g_{\omega K}\omega - \frac{g_{\rho K}}{2}\rho 
\end{eqnarray}

Using Eq.(5), the kaon energy $\omega_{K^{-}}$ can be calculated as a function of density. As the density increases, 
the electron chemical potential, $\mu_{e}$ also increases, and $\omega_{K^{-}}$ decreases. When $\omega_{K^{-}}$ decreases to a 
certain value such that $\mu_{K^{-}}$ = $\mu_{e}$, the $K^{-}$ starts appearing in the matter, a phase which is usually 
referred to as antikaon condensation.

The presence of $K^{-}$ modifies the field equations for $\sigma$, $\omega$, and $\rho$-mesons, 
which can be expressed as: 

\begin{eqnarray}
m_{\sigma}^{2}{\sigma} + g_{2}{\sigma}^{2} + g_{3}{\sigma}^{3} + U_{cut}^{'}(\sigma) 
= -g_{\sigma N}\left(n^{s}_{p} + n^{s}_{n}\right) -g_{\sigma K}n_{K}
\end{eqnarray}

\begin{eqnarray}
m_{\omega}^{2}\omega + c_{3}\omega^{3} + 2\Lambda_{v}g^{2}_{\omega N} g^{2}_{\rho N}\rho^{2}\omega 
= g_{\omega N}\left(n_{p} + n_{n}\right) -g_{\omega K}n_{K}
\end{eqnarray}

\begin{eqnarray}
m_{\rho}^{2}\rho + 2\Lambda_{v}g^{2}_{\omega N} g^{2}_{\rho N}\omega^{2}\rho
= \frac{g_{\rho N}}{2}\left(n_{p} - n_{n}\right) -\frac{g_{\sigma K}}{2}n_{K} 
\end{eqnarray}
Here, $n^{s}_{i}$, $n_{i}$, and 
$n_{K} = 2\left(\omega_{K^{-}} + g_{\omega K}\omega + \frac{g_{\sigma K}}{2}\rho\right)K^{*}K$
are the scalar density, number density of species $i$, and kaon density, respectively.  
The derivative of $U_{cut}(\sigma)$ \cite{ZHA.18} is given by:
\begin{eqnarray}
U_{cut}^{'}(\sigma) = \frac{\alpha\beta g_{\sigma N}}{M_{N}}
 \frac{1}{[ 1 + \exp\{-\beta(-g_{\sigma N}\sigma/M_{N}-f_{s})\}]}
\end{eqnarray}

The total energy density $\cal E$ of the charge neutral $\beta$-equilibrated neutron star matter 
with kaon condensation can be represented as
\begin{eqnarray}
{\cal E} &=& \sum_{i=p,n} \frac{1}{\pi^{2}}\int_{0}^{k_{Fi}}dk k^{2}\sqrt{k^{2} + M^{*2}_{N}}
+ \frac{1}{2}m^{2}_{\sigma}\sigma^{2} + \frac{1}{3}g_{2}\sigma^{3}\nonumber\\ 
&+& \frac{1}{4}g_{3}\sigma^{4} + U_{cut}(\sigma) + \frac{1}{2}m^{2}_{\omega}\omega^{2} + \frac{3}{4}c_{3}\omega^{4} 
+ \frac{1}{2}m^{2}_{\rho}\rho^{2}\nonumber\\ 
&+& 3\Lambda_{v}\left(g^{2}_{\omega N}\omega^{2}\right)\left(g^{2}_{\rho N}\rho^{2}\right) 
+ {\cal E}_{K^{-}} + \sum_{L} {\cal E}_{L} 
\label{eden}
\end{eqnarray}
and similarly the total pressure can be calculated as
\begin{eqnarray}
P &=& \sum_{i=p,n} \frac{1}{3\pi^{2}}\int_{0}^{k_{Fi}}dk k^{2}\frac{k^{2}}{\sqrt{k^{2} + M^{*2}_{N}}} 
- \frac{1}{2}m^{2}_{\sigma}\sigma^{2} - \frac{1}{3}g_{2}\sigma^{3}\nonumber\\
&-& \frac{1}{4}g_{3}\sigma^{4} - U_{cut}(\sigma) + \frac{1}{2}m^{2}_{\omega}\omega^{2} + \frac{1}{4}c_{3}\omega^{4}
+ \frac{1}{2}m^{2}_{\rho}\rho^{2}\nonumber\\ 
&+& \Lambda_{v}\left(g^{2}_{\omega N}\omega^{2}\right)\left(g^{2}_{\rho N}\rho^{2}\right) 
+ \sum_{L} {P}_{L} 
\label{pden}
\end{eqnarray}

In Eq. (\ref{eden}) ${\cal E}_{K^{-}}$ is the energy density contributed by kaon condensation and is given by:
\begin{eqnarray} 
{\cal E}_{K^{-}} = 2m^{*2}_{K}K^{*}K = m^{*2}_{K}n_{K}
\end{eqnarray}
Since the antikaon is an (s-wave) Bose condensate, it does not directly contribute to the pressure. 
However, the presence of the antikaon affects the fields, which in turn affects the pressure. 
The energy density and pressure from leptons (i.e., electrons and muons) 
are denoted by ${\cal E}_{L}$ and $P_{L}$, respectively. 

\subsection{Stellar Equations and Tidal Deformability}

The mass-radius relation for a neutron star is obtained by the Tolman, Oppenheimer, and Volkoff (TOV) equation 
\cite{OPP.39,TOL.39}, which is given by:
\begin{equation}
\frac{dP}{dr}=-\frac{G}{r}\frac{\left[\varepsilon+P\right ]
\left[M+4\pi r^3 P\right ]}{(r-2 GM)},
\label{tov1}
\end{equation}
\begin{equation}
\frac{dM}{dr}= 4\pi r^2 \varepsilon,
\label{tov2}
\end{equation}
\noindent
We use natural units, where c = 1, G, $P(r)$, and $M(r)$ represent the universal
gravitational constant, pressure of a neutron star, and the enclosed gravitational mass
inside a sphere of radius $(r)$ respectively.
Equations (\ref{tov1}) and (\ref{tov2}) are solved to determine the structural properties of a
static neutron star composed of charge-neutral matter \cite{LAT.04,KRA.06}. 

The tidal deformability parameter $\lambda$ is defined as \cite{FLA.08,HIN.08,HIN.10,DAM.12}:
\begin{equation}
Q_{ij}=-\lambda {\cal E}_{ij},
\end{equation}
where $Q_{ij}$ represents the induced {quadrupole} moment of a star in a binary system as a result of the static external 
tidal field ${\cal E}_{ij}$ from the companion star. The parameter $\lambda$ can be defined in relation to the 
dimensionless {quadrupole} tidal Love  number $k_2$ as:
\begin{equation}
\label{eq2}
\lambda = \frac{2}{3}k_2R^5,
\end{equation}
where $R$ represents the radius of the NS. The value of $k_2$ typically falls in 
the range of approximately 0.05 to 0.15 \cite{HIN.08,HIN.10,POS.10} for NSs and is dependent on the stellar structure. This quantity can be determined using the following expression \cite{HIN.08},
\bea
&& k_2 = \frac{8C^5}{5}\left(1-2C\right)^2
\left[2+2C\left(y_R-1\right)-y_R\right]\times\nonumber \\
&&\bigg\{2C\left[6-3 y_R+3 C(5y_R-8)\right]\nonumber\\
&&+4C^3\left[13-11y_R+C(3 y_R-2)+2
C^2(1+y_R)\right] \nonumber\\
&& ~ ~
+3(1-2C)^2\left[2-y_R+2C(y_R-1)\right]\log\left(1-2C\right)\bigg\}^{-1},
\eea
where $C$ $(\equiv M/R)$ is the compactness parameter of the star with 
mass $M$.  The value of $y_R$ $(\equiv y(R))$ can be found by solving
the following differential equation
\bea
r \frac{d y(r)}{dr} + {y(r)}^2 + y(r) F(r) + r^2 Q(r) = 0
\label{TidalLove2} ,
\eea
with
\bea
F(r) = \frac{r-4 \pi r^3 \left( \varepsilon(r) - P(r)\right) }{r-2
M(r)},
\eea
\bea
\nonumber Q(r) &=& \frac{4 \pi r \left(5 \varepsilon(r) +9 P(r) +
\frac{\varepsilon(r) + P(r)}{\partial P(r)/\partial
\varepsilon(r)} - \frac{6}{4 \pi r^2}\right)}{r-2M(r)} \\
&-&  4\left[\frac{M(r) + 4 \pi r^3
P(r)}{r^2\left(1-2M(r)/r\right)}\right]^2 \ .
\eea
{In the previous equations,} $M(r)$ represents the mass enclosed within the radius
$r$, while $\varepsilon(r)$ and $P(r)$ represent the energy density and pressure, respectively, in terms of 
the radial coordinate $r$ of a star.
These quantities are calculated within the chosen nuclear matter model to describe the stellar EoS. For a given EoS, Eq.(\ref{TidalLove2}) can be integrated together with the Tolman-Oppenheimer-Volkoff equations using the initial boundary conditions 
$y(0) = 2$, $P(0)\!=\!P_{c}$ and $M(0)\!=\!0$,{ where} $y(0)$, $P_c$ and $M(0)$ are 
the dimensionless quantity, pressure and mass at the center of the NS, respectively. 
The dimensionless tidal deformability can then be defined as $\Lambda = \frac{2}{3}k_2 C^{-5}$.

\section{Results and Discussion}
\begin{figure}
\includegraphics[width=8cm,height=5cm,angle=0]{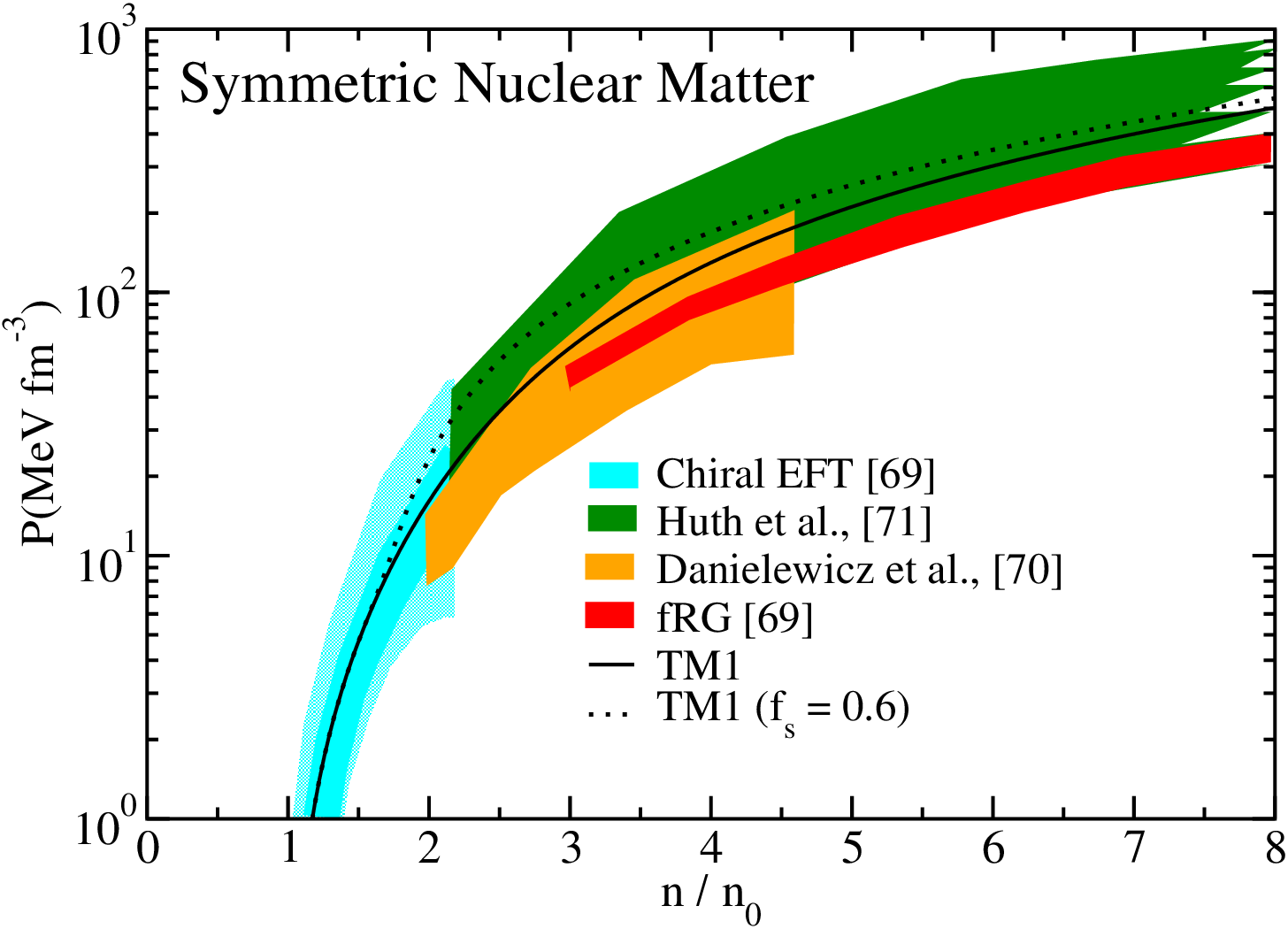}
\caption{The pressure of symmetric nuclear matter as a function of nucleon number density $n$ (in units of $n_{0}$) is shown for 
TM1 and TM1 with $f_{s}$ = 0.6, along with additional constraints (as mentioned in the text), represented by shaded regions.} 
\label{fig1}
\end{figure}

We now discuss the results of our investigation of neutron stars with antikaon condensation under TM1 and TM1e interactions 
with and without the $\sigma$-cut potential. The impact of the $\sigma$-cut potential on the underlying pressure can be seen 
from Fig. \ref{fig1}, where we show the pressure as a function of nucleon number density ($n$) (in units of $n_{0}$) for 
symmetric nuclear matter (SNM) for TM1. Alongside other constraints are also compared from various studies such as the 
one from the chiral EFT (cyan band) \cite{LEO.20}, ones derived from the heavy-ion collisions data \cite{DAN.02} 
(orange band), functional renormalization group (fRG) methods based on QCD by \cite{LEO.20} (red band), 
and recently proposed equations of state (EoSs) from \cite{HUT.21} (green band). 
It's worth noting that the pressure of symmetric nuclear matter for both TM1e and TM1 remains the same since 
there's no contribution from the $\rho$-meson in the symmetric nuclear matter. 
From Fig. 1, it's evident that both TM1 and TM1 with $f_{s}$ = 0.6 are in 
agreement with other constraints across all values of nucleon number density ($n$) in symmetric nuclear matter. However, both 
the equation of state, i.e., TM1 and TM1 with $f_{s}$ = 0.6 predict slightly higher pressure values compared to fRG (red band) 
within the density range of 3$\le$$n/n_{0}$$\le$10. The EoS with $f_{s}$ = 0.6 is not consistent with the prediction from the heavy-ion 
collision data for SNM.

\begin{figure}
\includegraphics[width=8cm,height=5cm,angle=0]{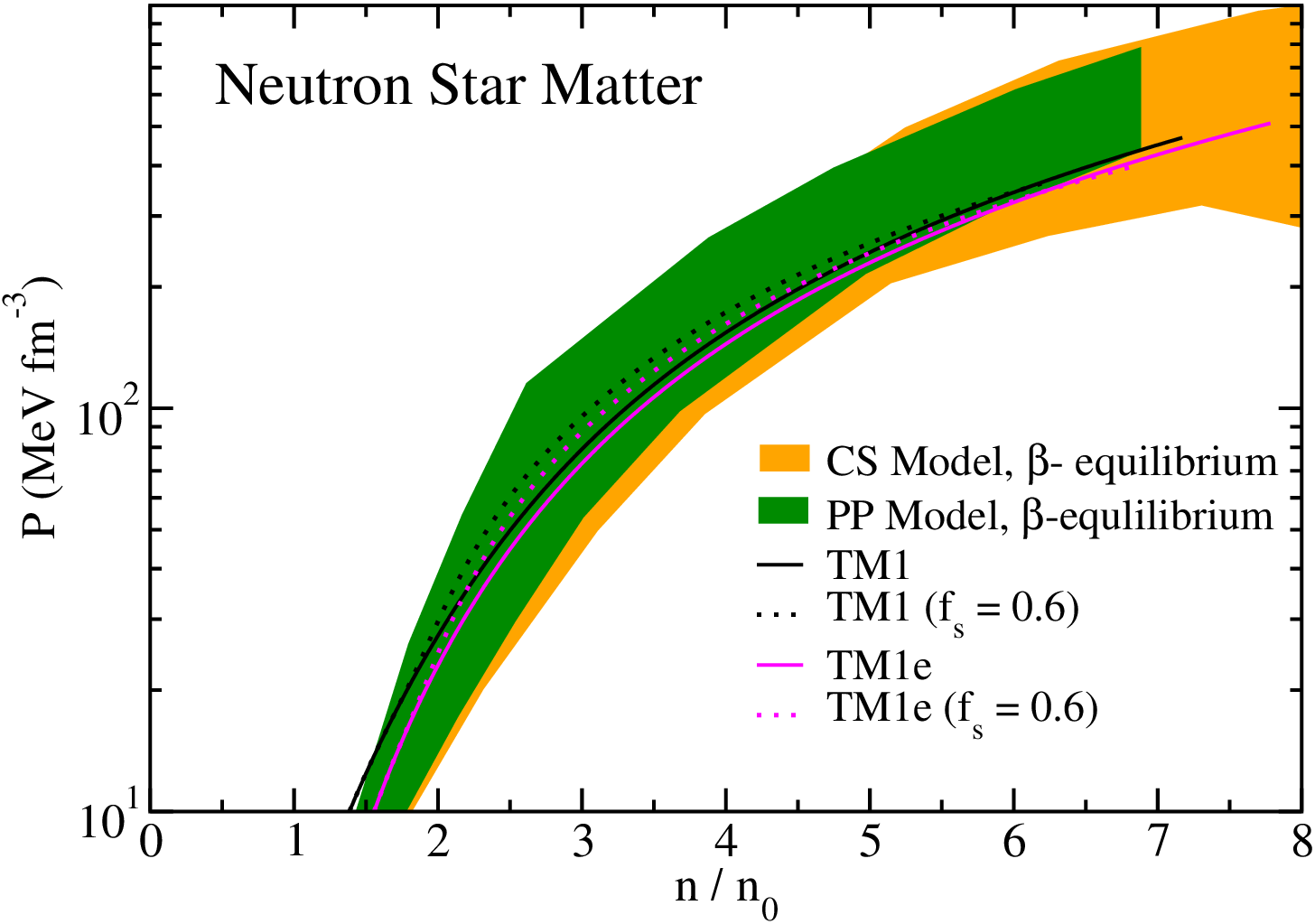}
\caption{The pressure of neutron star matter as a function of 
nucleon number density $n$ (in units of $n_{0}$) is shown 
for TM1 and TM1 with $f_{s}$ = 0.6, as well as for TM1e and TM1e with $f_{s}$ = 0.6, 
along with additional constraints represented by shaded regions.\\\\} 
\label{fig2}
\end{figure}

In Fig. \ref{fig2}, the calculated pressure of neutron star matter is plotted against the nucleon number density $n$ 
(in units of $n_{0}$) for TM1 and TM1 with $f_{s}$ = 0.6, alongside TM1e and TM1e with $f_{s}$ = 0.6. A recent
advancement in precise radius measurements was achieved by the Neutron Star Interior Composition
Explorer (NICER) collaboration \cite{RIL.19,MIL.19}, which determined the 
radius and mass of PSR J0030+0451 simultaneously through x-ray pulse-profile modeling.
Raaijmakers et al. \cite{RAA.19} investigated the implications of this measurement on the equation of
state (EoS) by utilizing two parametrizations for the neutron star EoS (in $\beta$-equilibrium): a piecewise
polytropic (PP) model with varying transition densities between the polytropes \cite{HEB.13} and 
a speed of sound (CS) model based on physical considerations at both nuclear and high densities \cite{GRE.19}.
Raaijmakers et al. \cite{RAA.20} conducted a combined analysis of these models to deduce implications on the
EoS from the NICER measurement, GW170817, and the 2.14 $M_{\odot}$. Their findings for the pressure as
a function of density are depicted i.e., CS model (orange band) and PP model (green band) in Fig. \ref{fig2}.
It is evident that TM1 and TM1e remain consistent with the CS and PP models. 
Introduction of the $\sigma$-cut potential $f_s$= 0.6 also maintains consistency 
with both CS and PP models across all values of nucleon number density ($n$). 

\begin{figure}
\includegraphics[width=9cm,height=6cm,angle=0]{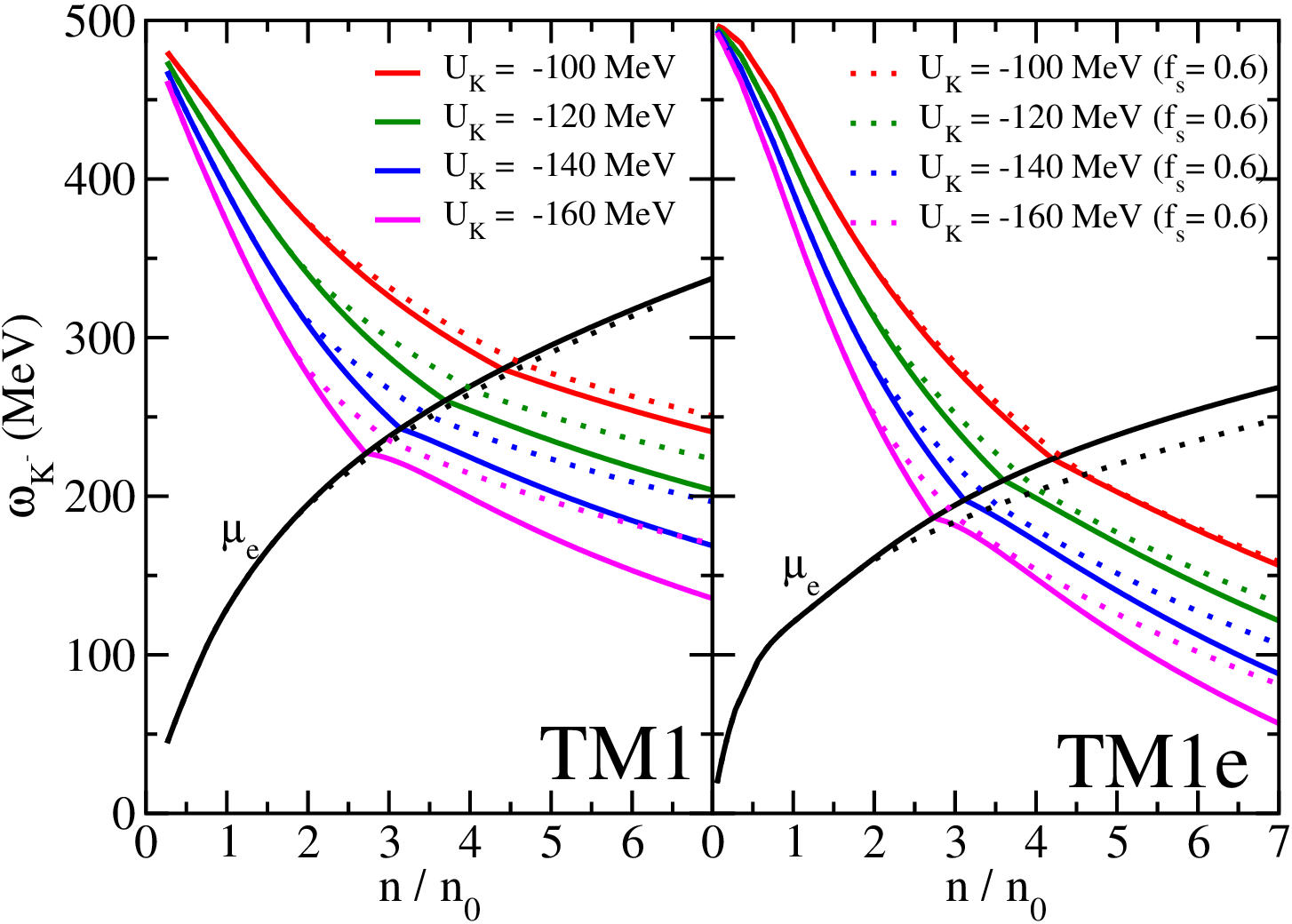}
\caption{The in-medium (anti)Kaon energy is plotted as a function of nucleon number density $n$ (in units of $n_{0}$) in a 
neutron star with the TM1 and TM1e parameter sets without (solid line) and with (dotted line) $f_{s}$ = 0.6 
for different kaon potentials as indicated in the left and right panels, respectively. The electron chemical potential
$\mu_{e}$ in the absence of kaons without (solid black line) and with (dotted black line) $f_{s}$ = 0.6  
is shown in both panels.} 
\label{fig3}
\end{figure}

The left and right panels of Figure \ref{fig3} display the in-medium antikaon energy ($\omega_{K^{-}}$) plotted against nucleon 
number density, normalized to $n_{0}$ for the models TM1 and TM1e respectively. 
In the left panel of figure \ref{fig3}, the solid lines shows the results of TM1 equation of state (EoS), while the dotted lines
illustrates the outcomes with the inclusion of a $\sigma$-cut potential with $f_s$ = 0.6. From both the panels, it is evident that 
the energy of antikaons energy decreases with density as expected. Moreover, the curve of the electron chemical potential  
($\mu_e$) intersects the
kaon energy at different densities, corresponding to different (anti)kaon potentials, marking the end of the pure hadronic phase
and the onset of the condensate phase. 
The dotted lines in the figure illustrate the effect of the $\sigma$-cut potential on the in-medium antikaon energy,
which becomes noticeable around 2$n_{0}$ for all values of $U_{K^{-}}$. At higher densities from 2$n_{0}$ to 7$n_{0}$,
the impact of the $\sigma$-cut potential becomes more significant for deeper values of $U_{K^{-}}$  
with highest effect observed for $U_{K^{-}}$ = -160 MeV and the lowest for $U_{K^{-}}$ = -100 MeV. 
The $\sigma$-cut potential is known to reduce the influence of the $\sigma$ field, 
resulting in a stiffer antikaon energy ($\omega_{K^{-}}$) at high densities, 
it is more pronounced for $U_{K^{-}}$ = -160 MeV due to the higher contribution of the $\sigma$-field  
compared to other $U_{K^{-}}$ values.     

Similarly, solid lines represent the results of the TM1e equation of state (EoS), while the dotted lines
illustrate the outcomes with the inclusion of a $\sigma$-cut potential with $f_s$ = 0.6 in the right panel of Figure \ref{fig3}. 
In this case, the antikaon energy decreases with density for all values of $U_{K^{-}}$, and it reduces
to a lower value as the density increases for a given value of $U_{K^{-}}$ due to the 
additional $\Lambda_{v}$ coupling as well as a larger value of $g_{\rho}$ in TM1e compared to TM1. 
The overall effect of the $\sigma$-cut potential
is more or less the same as TM1. 
However, the most significant effect of the $\sigma$-cut potential
and symmetry energy is on the electron chemical potential ($\mu_{e}$) in TM1e compared to TM1 with and without $f_{s}$. 
The electron chemical potential 
$\mu_{e}$ is more influenced by the isovector channel, specifically the contribution of the $\rho$-field compared to  
the $\sigma$ and $\omega$- fields. As a result, TM1e, with a higher $g_{\rho}$ value and additional $\Lambda_{v}$ coupling 
leads to a significant decrease in the value of $\mu_{e}$ after 2$n_{0}$ for TM1e with $f_{s}$ = 0.6 compared to TM1 
with $f_{s}$ = 0.6.

\begin{figure}
\includegraphics[width=8cm,height=5cm,angle=0]{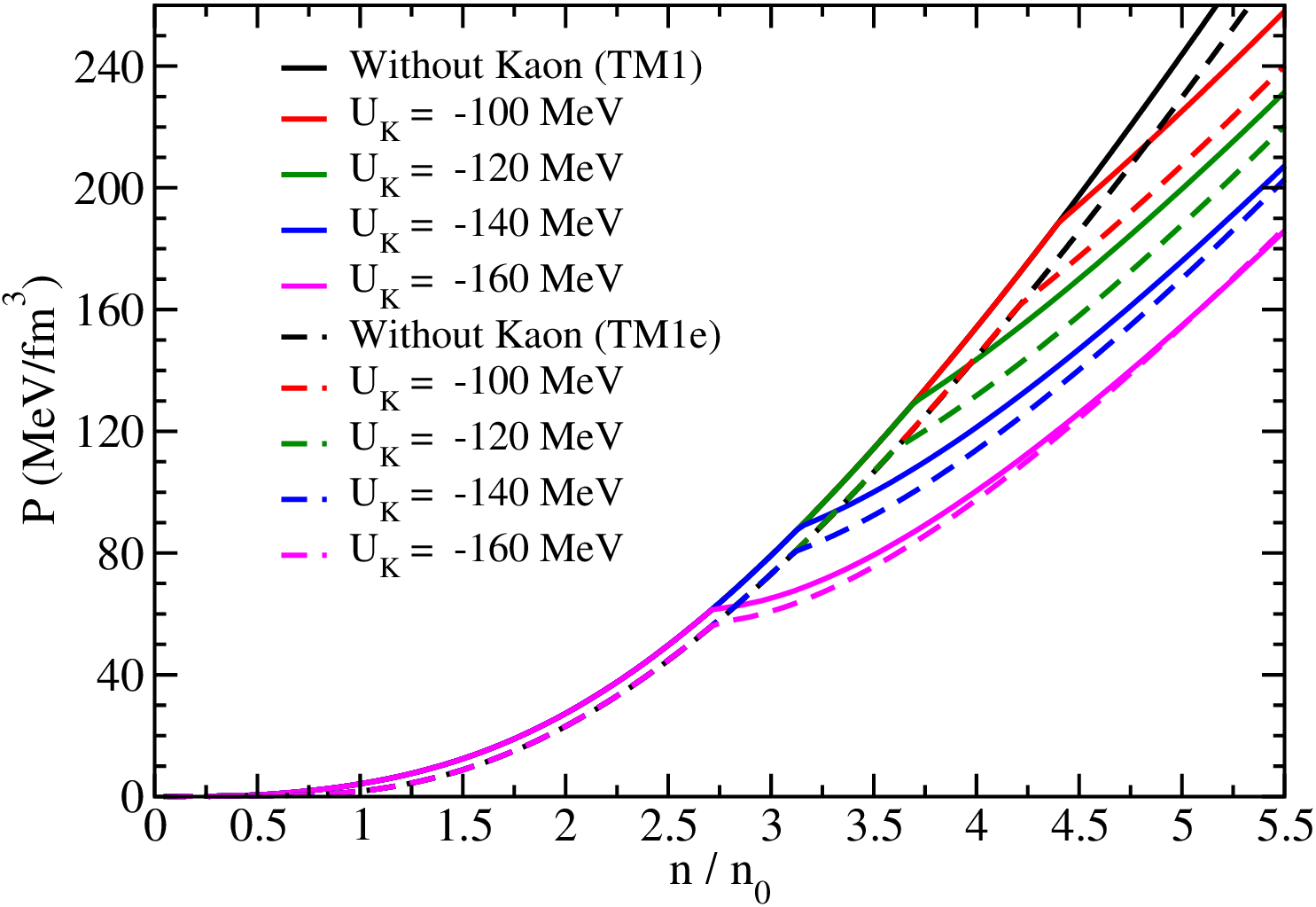} 
\caption{Pressure (P) as a function of normalized nucleon density is plotted for TM1 (solid line)
and TM1e (dashed line). The scenario without $K^-$ (black line) 
is compared with pressure curves including $K^-$ condensates for different potentials such as $U_{K^{-}}$= -100 MeV (red), 
$U_{K^{-}}$= -120 MeV (green), $U_{K^{-}}$= -140 MeV (blue), and $U_{K^{-}}$= -160 MeV (magenta).\\\\}  
\label{fig4}
\end{figure}

In Fig. \ref{fig4}, we illustrate the equation of state (EoS) by showing the variations of pressure with respect to 
nucleon density normalized to $n_{0}$
for different values of (anti)kaon potential $U_{K^{-}}$ for TM1 (solid line) and TM1e (dashed line). 
The TM1e, an extension of TM1 with $\Lambda_{v}$ coupling, results in a softer EoS
at high density for all $U_{K^{-}}$ values. In TM1e, antikaons appear at
lower densities compared to TM1, with this difference being more pronounced for 
$U_{K^{-}}$ values of -100 and -120 MeV. The nucleon density difference at which   
antikaons occur between TM1 and TM1e decreases for deeper $U_{K^{-}}$ values.
The phase transition of antikaons is second order for all $U_{K^{-}}$ values 
in both TM1 and TM1e parameter sets. 

\begin{figure}
\includegraphics[width=8cm,height=5cm,angle=0]{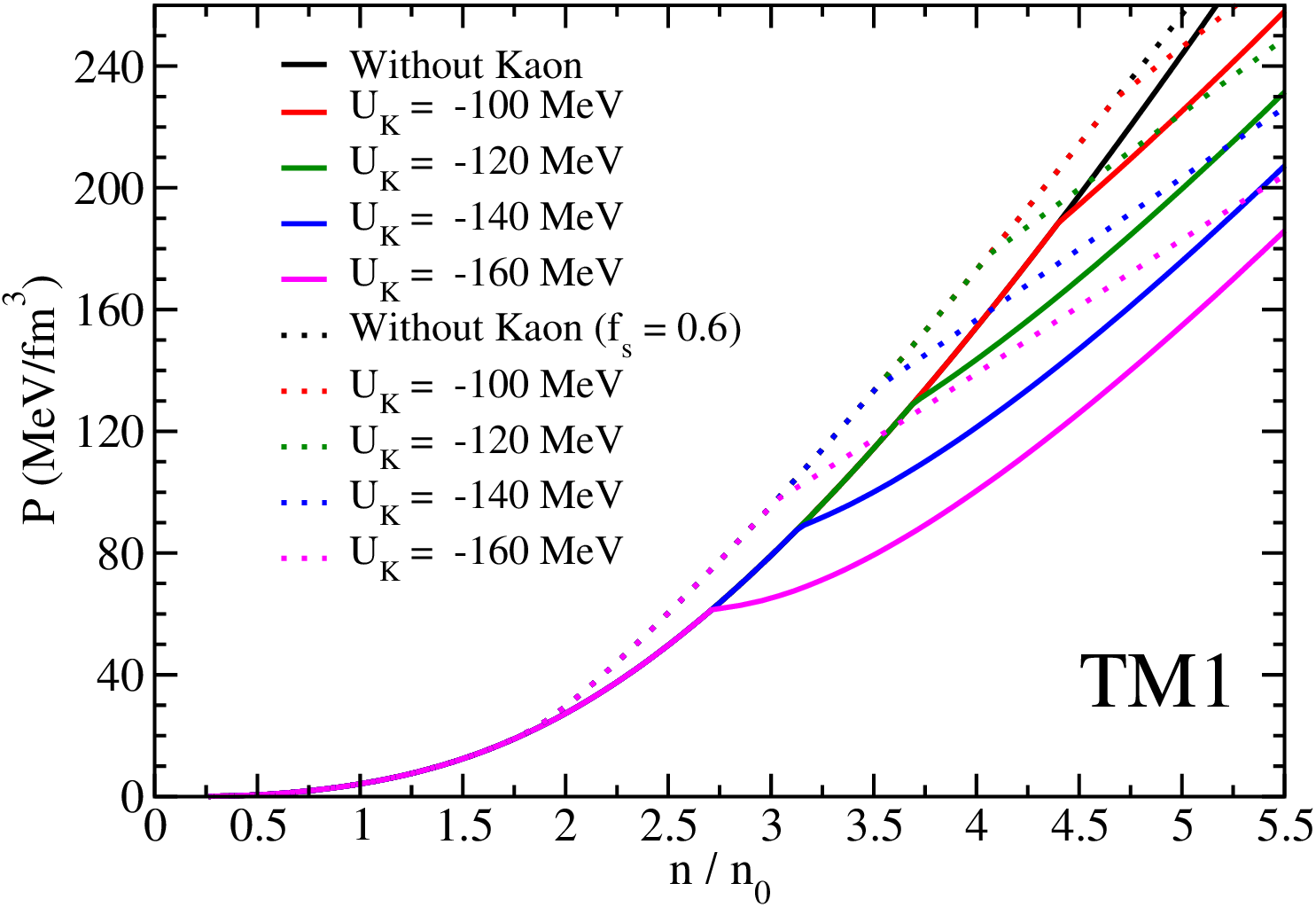} 
\caption{Pressure (P) as a function of normalized nucleon density is plotted for TM1. The scenario without 
$K^-$ (black line) 
is compared with pressure curves including $K^-$ condensates for different potentials such as $U_{K^{-}}$= -100 MeV (red), 
$U_{K^{-}}$= -120 MeV (green), $U_{K^{-}}$= -140 MeV (blue), and $U_{K^{-}}$= -160 MeV (magenta). The dotted line shows the 
results by incorporating the $\sigma$-cut potential with $f_s$ = 0.6.}
\label{fig5}
\end{figure}

\begin{figure}
\includegraphics[width=7cm,height=5cm,angle=0]{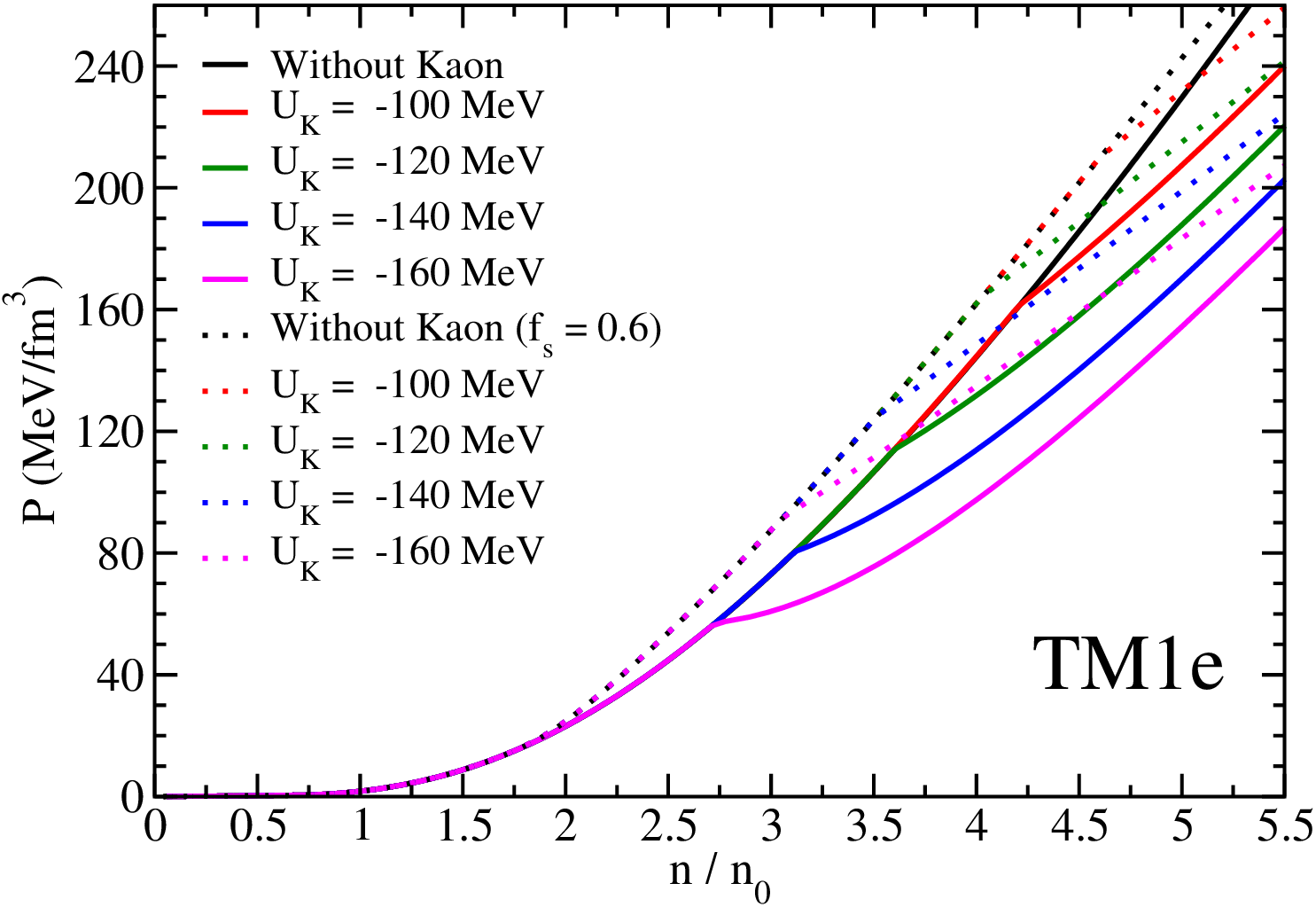} 
\caption{Same as Fig. \ref{fig5}, but for TM1e.} 
\label{fig6}
\end{figure}

Figures \ref{fig5} and \ref{fig6} illustrate the impact of the (anti)kaon potential on the pressure as 
a function of normalized baryon density for TM1 and TM1e, respectively. The solid lines
in both figures represent the effect of the (anti)kaon potential on the pressure for 
the TM1 and TM1e parameter sets, while the dotted line in both figures shows the TM1 and TM1e with 
$f_s$ = 0.6, respectively. It is evident from Figure \ref{fig5} that the EoS becomes stiffer for TM1 
with $f_s$ = 0.6 compared to the TM1 parameter set. Similiarly, EoS of TM1e with $f_s$ = 0.6 is
stiffer compared to TM1e parameter set as shown in Figure \ref{fig6}. 
The TM1e model exhibits a softer profile compared to the TM1 model due to the 
$\Lambda_{v}$ coupling. The effect of antikaons remains consistent with TM1, 
with the softest behavior observed for $U_{K^{-}}$ = -160 MeV. 

When incorporating a $\sigma$-cut potential with $f_s$= 0.6, 
the EoS shows slightly increased stiffness for specific $U_{K^{-}}$ values compared to the 
counterpart as evident from both figures. The points at which the kaons begin to appear in dense matter are reflected 
as branching points on the pressure curve at those densities. The inclusion of the $\sigma$-cut potential delays the 
onset of the condensate phase, resulting in slightly stiffer pressure curves. It is important to note
that from Figures \ref{fig5} and \ref{fig6}, antikaon condensation is of second order for all the values of   
$U_{K^{-}}$ for both TM1 and TM1e with and without $\sigma$-cut potential.

\begin{figure}
\includegraphics[width=8cm,height=6cm,angle=0]{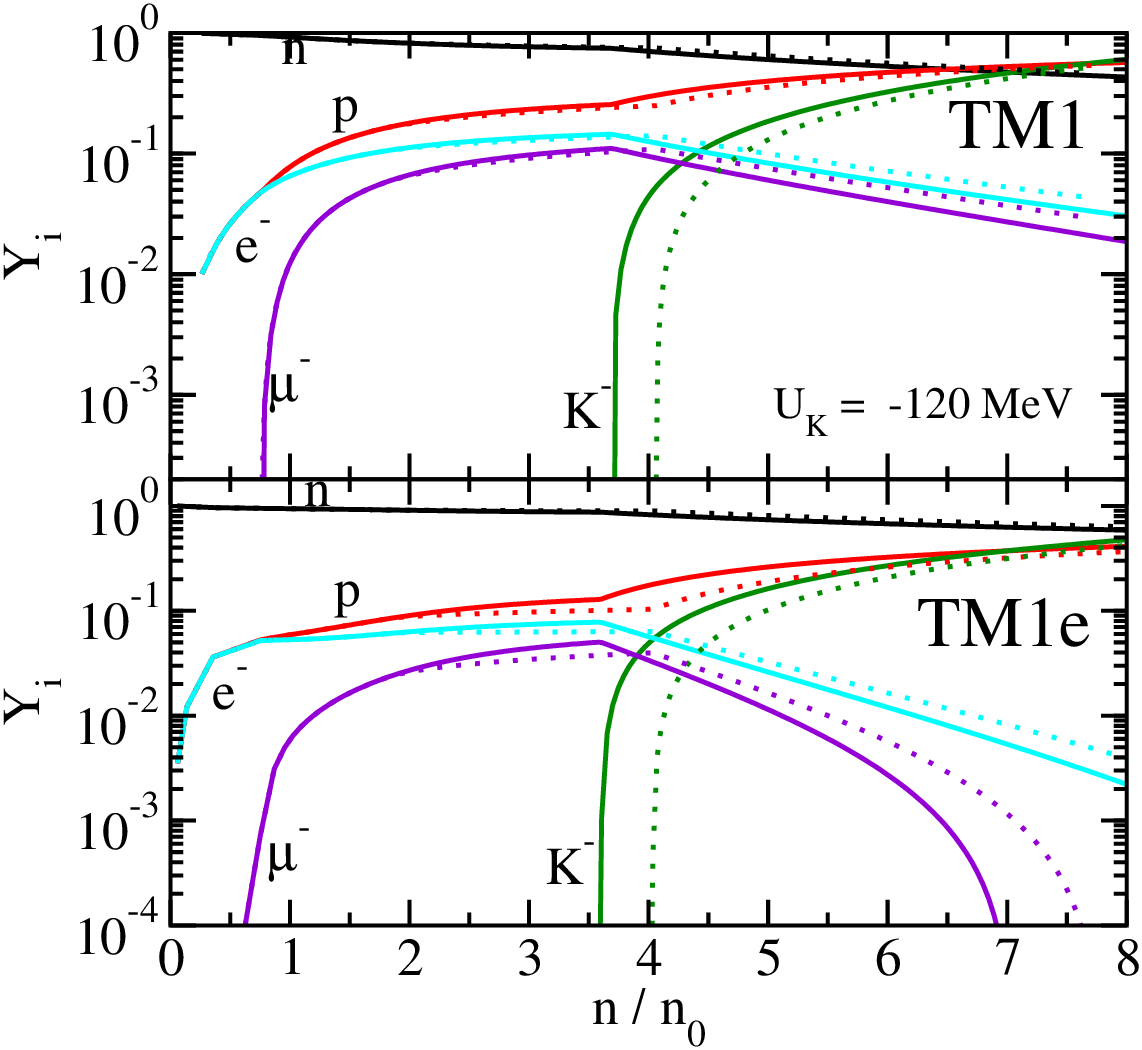} 
\caption{Fraction of various particles in $\beta$-equilibrated neutron, proton, lepton matter including $K^-$ condensates for 
$U_{K^{-}}$= -120 MeV as a function of normalized nucleon number density for TM1 (upper panel) and TM1e (lower panel). 
The associated dotted curves represent the fractions for $f_s$= 0.6.\\\\} 
\label{fig7}
\end{figure}

In Fig. \ref{fig7}, we show the particle fractions at a specific potential value of $U_{K^{-}}$ = -120 MeV, 
separately for 
TM1 (upper panel) and TM1e (lower panel). The associated dotted lines represents the corresponding particle fractions, 
when the $\sigma$-cut potential is incorporated for the two models. The analogous feature that is seen in both models is 
the density at which the (anti)kaons start appearing at roughly 3.7$n_0$. A more or less similar delay is seen in their appearance 
when the $\sigma$-cut potential is incorporated. Before the emergence of $K^{-}$, the system upholds charge neutrality among protons, 
electrons, and muons. It becomes evident that upon the onset of $K^-$ condensation, the system swiftly restores its 
charge neutrality, 
occurring at $\approx$ 3.7 $n_0$ for TM1 and 3.6 $n_0$ for TM1e, resulting in the subsequent deleptonization.
Therefore, the symmetry energy has a minor impact on the emergence of $K^{-}$.   
However, the deleptonization happens faster in the case of TM1e in comparison. This behavior is driven by the nuclear
symmetry energy, which is controlled by $g_{\rho}$ and $\Lambda_{v}$ coupling. When the antikaon condensation occurs, it increases 
with density, 
replacing leptons to maintain charge neutrality. This lowers the electron chemical potential $\mu_{e}$, leading to faster 
deleptonization in the case of TM1e due to a larger value of $g_{\rho}$ with additional $\Lambda_{v}$ coupling.    
This outcome aligns with the expectation that as $K^-$ mesons, being bosons, preferentially 
condense in the lowest-energy state, maintaining charge neutrality. As a result, there is an increase in the proton fraction, 
consequently leading to an almost isospin symmetric state at higher densities. 

\begin{figure}
\includegraphics[width=7.5cm,height=5cm,angle=0]{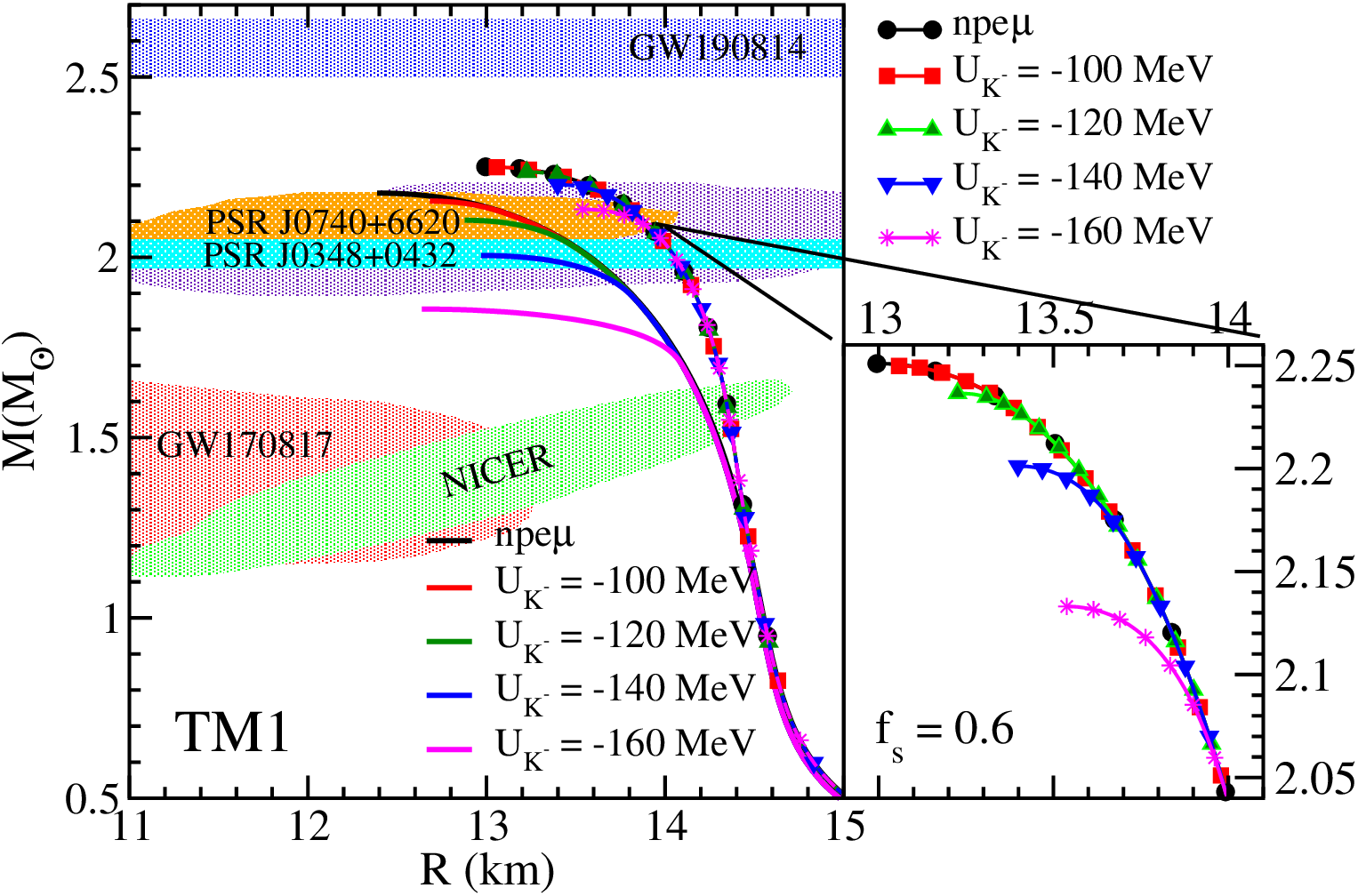} 
\caption{The MR solution curve of the neutron star with the TM1 model. Here the solid lines represent calculations without  
$\sigma$-cut, while the lines with symbols are the results with $\sigma$-cut ($f_s$ = 0.6) and is magnified in the adjacent figure. 
Black lines correspond to neutron, proton, and lepton matter, whereas other lines depict $K^-$ condensates computed for 
different $U_{K^{-}}$ potentials. The shaded region illustrates the available astrophysical constraints\label{fig8}.} 
\end{figure}

\begin{figure}
\includegraphics[width=7cm,height=5cm,angle=0]{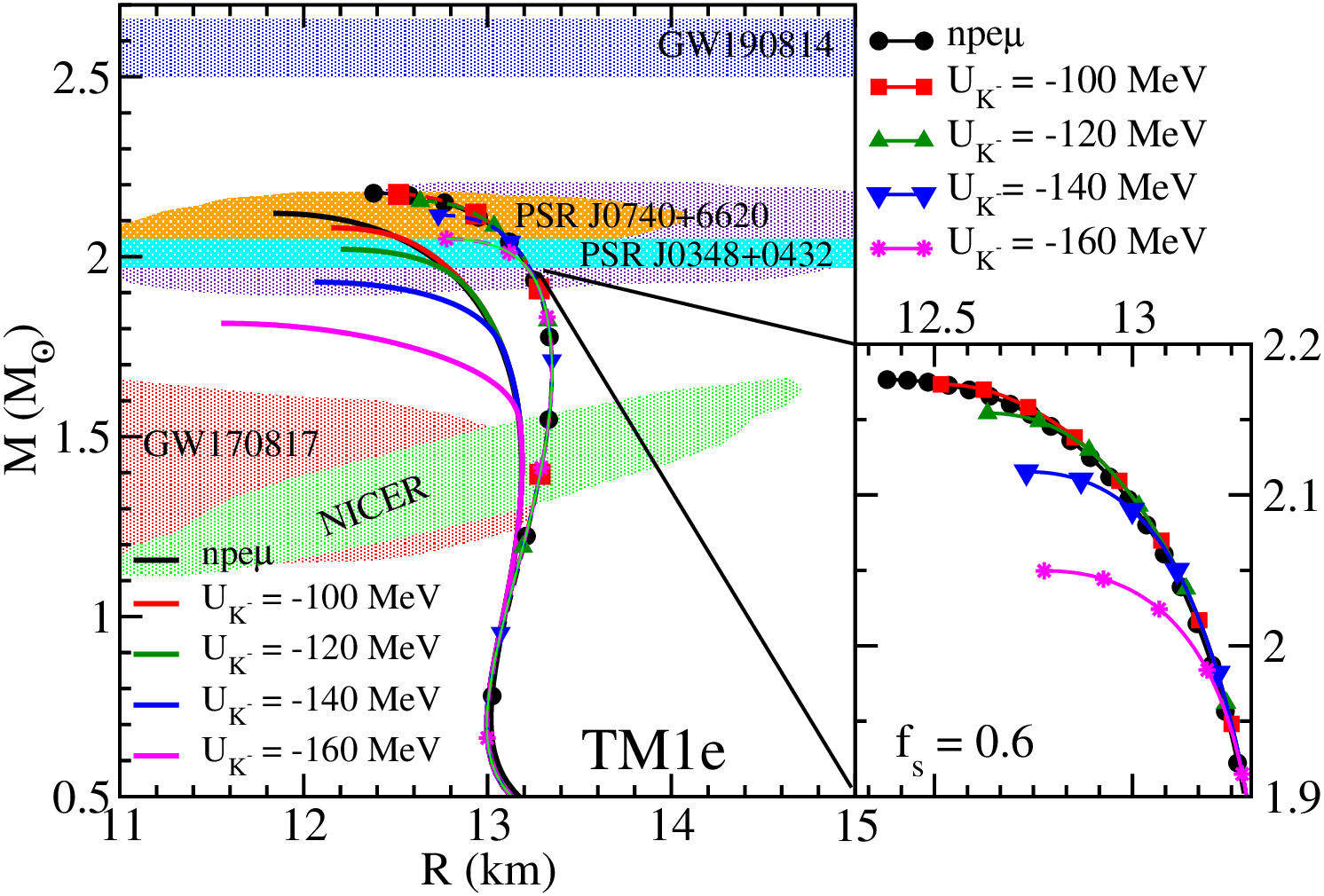} 
\caption{Same as Fig. \ref{fig8} but for TM1e. 
\label{fig9}} 
\end{figure}

The mass-radius relationship for neutron stars (NS) for the two models, TM1 and TM1e, is shown in Fig. \ref{fig8} and 
Fig. \ref{fig9}, with the right panel magnifying the results obtained by incorporating the $\sigma$-cut ($f_s = 0.6$) 
scheme with different (anti)kaon potentials. Black lines depict the baseline calculations involving neutron, proton, and lepton matter, 
whereas colored lines represent the effects of $K^-$ condensates with different optical potentials. The red shaded region indicates 
the astrophysical constraints derived from the LIGO GW170817 event \cite{ABB.17}. Additionally, the two-dimensional 
posterior distribution in the mass-radius domain obtained from NICER X-ray data for the millisecond pulsar PSR J0030+0451 
is shown. Other shaded regions represent constraints from PSR J0348+0432 (cyan) \cite{ANT.13} and PSR J0740+6620 \cite{FON.21,RIL.21}. 

The upper and lower panels of Table III show the maximum mass, radius, and central density for different $U_{K^{-}}$ values
for TM1 and TM1 with $f_{s}$= 0.6, respectively. Similarly, the upper and lower panels of Table IV provide the maximum mass, 
radius, and central density for different $U_{K^{-}}$ values for TM1e and TM1e with $f_{s}$= 0.6, respectively.  
We observed that the maximum mass decreases with an increase in the depth of the (anti)kaon potential for both models 
(with or without $\sigma-$cut), as shown in Tables III and IV, respectively.  
Neutron stars composed of only nucleons and leptons (without antikaon)
have the maximum masses of 2.178 $M_{\odot}$ and 2.251 $M_{\odot}$ for TM1 and TM1 with $f_{s}$= 0.6 and 
2.120 $M_{\odot}$ and 2.176 $M_{\odot}$ for TM1e and TM1e with $f_{s}$= 0.6, respectively.   
Notably, neutron stars with antikaon remain consistent with the observed maximum mass neutron star constraint of
approximately 2$M_{\odot}$ except -160 MeV for TM1 and except -140 and -160 MeV for TM1e, respectively.
On the other hand, the radius corresponding to the maximum mass for all the values of $U_{K^{-}}$ is below
13.0 and 13.6 km for TM1 and TM1 with $f_{s}$ = 0.6 as given in Table III, respectively. Similarly, one can find
from Table IV the radius corresponding to the maximum mass for all the values of $U_{K^{-}}$ is below
12.3 and 12.8 km for TM1e and TM1e with $f_{s}$ = 0.6, respectively.   
The equation of state (EoS) of TM1 and TM1e with $f_{s}$= 0.6 for all values of $U_{K^{-}}$ satisfies
the observed maximum mass constraint of 2$M_{\odot}$. Overall, for a given depth of the (anti)kaon potential,
TM1 and TM1 with $f_{s}$= 0.6 have a slightly higher maximum mass and radius compared to TM1e and TM1e with $f_{s}$= 0.6.
This is because TM1e has a $\Lambda_{v}$ coupling, which makes the EoS softer at high densities.

\begin{table}
\centering
\caption{The maximum mass, $M_{max}$ (in units of $M_\odot$), radius (in km), corresponding central density (in units of $n_{0}$) 
of neutron stars for different values of antikaon optical potential depths $U_{{K^{-}}}$ (in units of MeV) at
$n_{0}$ for TM1 (upper panel) and TM1 with $f_{s}$ = 0.6 (lower panel).}
\begin{tabular}{ccccc}
$U_{{K^{-}}}$ (MeV) & $M_{max}$ ($M_\odot$) & R (km) & $n_{c}(n_{0})$ \\
\hline
\hline
0    & 2.178 & 12.388 & 5.884 \\
-100 & 2.156 & 12.683 & 5.641 \\
-120 & 2.103 & 12.878 & 5.436 \\
-140 & 2.005 & 12.970 & 5.354 \\
-160 & 1.856 & 12.638 & 5.839 \\
\hline
0    & 2.251 & 12.995 & 5.044 \\
-100 & 2.250 & 13.058 & 5.051 \\
-120 & 2.237 & 13.226 & 4.908 \\
-140 & 2.201 & 13.398 & 4.726 \\
-160 & 2.133 & 13.538 & 4.566 \\
\hline
\hline
\end{tabular}
\end{table}

\begin{table}
\centering
\caption{The maximum mass, $M_{max}$ (in units of $M_\odot$), radius (in km), corresponding central density (in units of $n_{0}$) 
of neutron stars for different values of antikaon optical potential depths $U_{{K^{-}}}$ (in units of MeV) at
$n_{0}$ for TM1e (upper panel) and TM1e with $f_{s}$ = 0.6 (lower panel).}
\begin{tabular}{ccccc}
\hline
\hline
$U_{{K^{-}}}$ (MeV) & $M_{max}$ ($M_\odot$) & R (km) & $n_{c}(n_{0})$ \\
\hline
0    & 2.120 & 11.836 & 6.273 \\
-100 & 2.080 & 12.151 & 5.951 \\
-120 & 2.021 & 12.202 & 5.889 \\
-140 & 1.929 & 12.060 & 6.112 \\
-160 & 1.815 & 11.553 & 6.951 \\
\hline
0    & 2.176 & 12.380 & 5.420 \\
-100 & 2.173 & 12.516 & 5.335 \\
-120 & 2.154 & 12.634 & 5.196 \\
-140 & 2.115 & 12.732 & 5.065 \\
-160 & 2.050 & 12.777 & 4.982 \\
\hline
\hline
\end{tabular}
\end{table}

\begin{figure}
\includegraphics[width=7.5cm,height=5.5cm,angle=0]{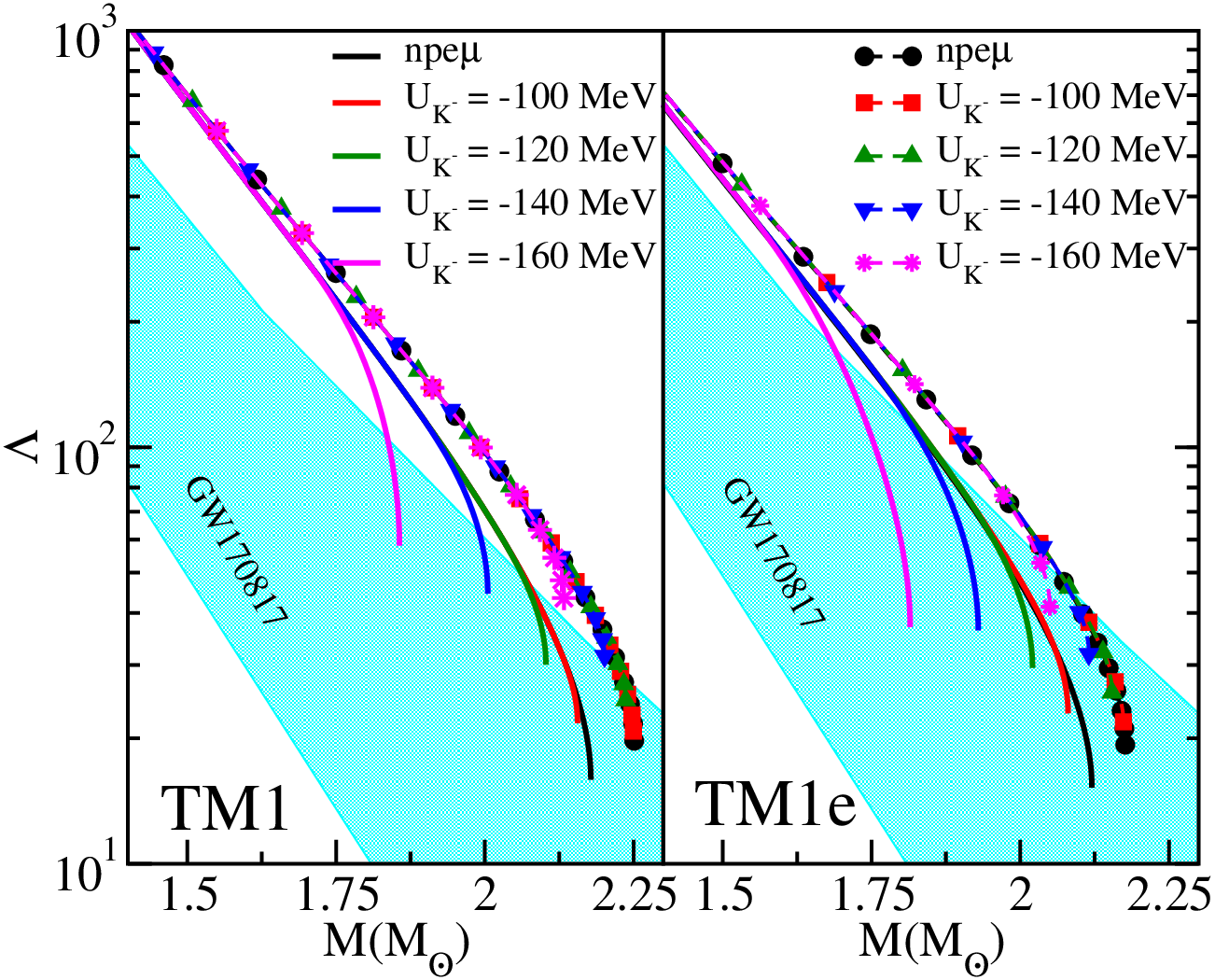} 
\caption{ The left panel displays the tidal deformability ($\Lambda$) of a neutron star as a function of mass for the TM1 model, 
while the right panel shows the predictions of the TM1e model. Solid lines in both panels represent calculations without 
the $\sigma$-cut, while those with symbols are with a $\sigma$-cut ($f_s$ = 0.6) scheme. The neutron, proton, and lepton matter 
are represented by black lines, whereas others denote $K^-$ condensates for different $U_{K^{-}}$ potentials. 
The shaded region indicates the available astrophysical constraints.} 
\label{fig10}
\end{figure}

\begin{table}
\centering
\caption{The threshold densities, $n_{cr}$ (in units of $n_{0}$) for antikaon condensation in dense nuclear
matter for different values of $U_{{K^{-}}}$ (in units of MeV) at $n_{0}$. The corresponding mass 
$M$ (in units of $M_\odot$) and radius (in km) of neutron stars for TM1 (upper panel) and TM1 with
$f_{s}$ = 0.6 (lower panel).}
\begin{tabular}{ccccc}
\hline
\hline
$U_{{K^{-}}}$ (MeV)& $n_{cr}(n_{0})$ & $M$ ($M_\odot$) & R (km)\\
\hline
-100 & 4.436 & 2.125 & 13.090 \\
-120 & 3.725 & 2.025 & 13.504 \\
-140 & 3.160 & 1.875 & 13.847 \\
-160 & 2.850 & 1.705 & 14.091 \\
\hline
-100 & 4.692 & 2.248 & 13.155 \\
-120 & 4.064 & 2.219 & 13.463 \\
-140 & 3.518 & 2.151 & 13.756 \\
-160 & 3.032 & 2.026 & 14.018 \\
\hline
\hline
\end{tabular}
\end{table}

\begin{table}
\centering
\caption{The threshold densities, $n_{cr}$ (in units of $n_{0}$) for antikaon condensation in dense nuclear
matter for different values of $U_{{K^{-}}}$ (in units of MeV) at $n_{0}$. The corresponding mass 
$M$ (in units of $M_\odot$) and radius (in km) of neutron stars for TM1e (upper panel) and TM1e with
$f_{s}$ = 0.6 (lower panel).}
\begin{tabular}{ccccc}
\hline
\hline
$U_{{K^{-}}}$ (MeV)& $n_{cr}(n_{0})$ & $M$ ($M_\odot$) & R (km)\\
\hline
-100 & 4.213 & 2.013 & 12.674 \\
-120 & 3.552 & 1.874 & 12.951 \\
-140 & 3.123 & 1.723 & 13.100 \\
-160 & 2.776 & 1.532 & 13.178 \\
\hline
-100 & 4.630 & 2.164 & 12.694 \\
-120 & 4.037 & 2.118 & 12.937 \\
-140 & 3.523 & 2.033 & 13.144 \\
-160 & 3.077 & 1.894 & 13.293 \\
\hline
\hline
\end{tabular}
\end{table}

The threshold densities $n_{cr}$ for antikaon condensation in dense
nuclear matter for different values of $U_{K^{-}}$ along with corresponding mass and radius
for TM1 and TM1 with $f_{s}$ are shown in Table V. Similar results for TM1e and TM1e with $f_{s}$
are given in Table VI. One can see from the observation of Tables V and VI that with an increase in the depth of 
the (anti)kaon potential, the threshold density $n_{cr}$ for antikaon condensation and the corresponding
mass decrease while the radius is increasing. This trend is consistent  regardless of the parameter
sets used. From Tables V and VI, one can observe that antikaon appears at the lowest
density for $U_{K^{-}}$ = -160 MeV irrespective of the parameter sets.  
It is to be noted that there is a slight variation in the value of 
$n_{cr}$ and corresponding mass and radius for different values of $U_{K^{-}}$ for TM1 and TM1e as well as
TM1 and TM1e with $f_{s}$. For a given value of $U_{K^{-}}$, the TM1e and TM1e with $f_{s}$ predict
a smaller value of mass compared to TM1 and TM1 with $f_{s}$. This is due to the $\Lambda_{v}$ coupling
which makes the EoS softer at high densities. From Table V, we find that the mass corresponding to $n_{cr}$
is above 2$M_{\odot}$ for $U_{K^{-}}$ = -100 and -120 MeV for TM1 (upper panel).
But for TM1 with $f_{s}$ = 0.6 (lower panel), the mass corresponding to $n_{cr}$ is above 2$M_{\odot}$ for
all values of $U_{K^{-}}$. Similarly, as TM1e has a $\Lambda_{v}$ coupling, only $U_{K^{-}}$ = -100 MeV
predicts the mass approximately 2$M_{\odot}$ and TM1e with $f_{s}$ = 0.6 (lower panel) the mass 
corresponding to $n_{cr}$ is above 2$M_{\odot}$ except $U_{K^{-}}$ = -160 MeV.
The $\sigma$-cut potential has an impact on the mass corresponding to $n_{cr}$, but the variation in
$n_{cr}$, corresponding mass, and radius for different values of $U_{K^{-}}$ is not significant. 

In Fig. \ref{fig10}, we present the results obtained for the tidal deformability of the models considered here. 
It is known that in the final stages of the coalescence, neutron stars develop a mass quadrupole due to the extremely 
strong tidal gravitational field induced by the counterpart comprising the binary. The dimensionless tidal deformability
describes the degree of deformation of a neutron star and depends on the nature of the equation of state (EoS) 
\cite{HIN.10,DAM.09,BIN.09,MAL.18}.
In this regard, several bounds on the tidal deformability parameter `$\Lambda_{1.4}$' (at mass $1.4 M_{\odot}$) have 
been calculated lately from different wavelength analyses of the GW170817 data. For example there is the lower limit 
$\Lambda_{1.4} > 344$ \cite{RAD.18,PER.17} as well as the upper limits of $\Lambda_{1.4} < 800$ \cite{ABB.17}. 
From the plot in Fig. \ref{fig10}, overall we can 
see that both models, TM1 and TM1e, agree with the shaded area of the GW analysis over $\Lambda$ vs Mass, more so when condensates 
are not considered. However, TM1e results with (anti)kaon condensate at various potentials seem to be in better agreement 
with the data than that obtained with TM1. The impact of the condensate appears much later for the two models, and therefore, for 
all the schemes adopted in this work, we obtain $\Lambda_{1.4} =$ 900 and 690 for TM1 and TM1e, respectively, which is more or 
less consistent with the upper and lower bounds discussed earlier. Overall results from TM1e seem to be quite impressive 
with or without invoking the $\sigma$-cut scheme at all (anti)kaon potentials. The inclusion of the condensates, however, seems 
to lower the tidal deformability with an increase in mass.  

\section{Conclusions and Summary}
We adopt the $\sigma$-cut scheme, inevitably meant to stiffen the underlying EoS, thereby increasing the maximum mass obtained 
for neutron stars compared to well-known RMF models, TM1 and TM1e and investigate the effect of the same for neutron stars 
with (anti)kaon condensate obtained with different (anti)kaon optical potentials. The main difference between TM1 and TM1e 
is the density dependence of symmetry energy. From our present analysis, we found that TM1 and TM1e as well as TM1 and TM1e
with $\sigma$-cut potential are consistent with a speed of sound (CS) model and a piecewise polytropic (PP) model.  
The role of the symmetry energy associated with the two models is highlighted with the global properties of the 
neutron stars such as maximum mass, radius, composition and tidal deformability and compared with the constraints
imposed from other theoretical studies and observational data. 
The effect of the $\sigma$-cut scheme on the EoS is found to appear at $\approx$ 2$n_{0}$ and increases 
thereafter. The $K^{-}$ condensates start to appear at about 3.5$n_{0}$, contributing to the overall charge neutrality of the 
matter and dominating the population of other species thereby leading to deleptonization in matter as well. The appearance of 
the condensate is set earlier if the potential is deeper. We could obtain neutron star mass $M > 2 M_{\odot}$ with the scheme 
for most of the $K^{-}$ potentials considered here, which agrees with the recent observation of high mass stars such as 
$PSR J0740+6620$ and $PSR J0348+0432$. The calculated tidal deformability parameter is also found
be in agreement with GW data analysis, particularly for the TM1e case. 

In the present analysis, we set the value of the free parameter $f_{s}$ to maintain the properties of finite nuclei 
in the TM1 model without being influenced by the  $\sigma$-cut potential.
Our analysis highlights the significance of the $\sigma$-cut scheme in achieving a maximum mass for neutron stars above
2$M_{\odot}$, but it tends to over-estimate the corresponding tidal deformability of neutron stars. To address this 
discrepancy and incorporate constraints from other observational data related to the density dependence of symmetry, 
such as the low tidal deformability of neutron stars, it is necessary to include the $\omega$-$\rho$ ($\Lambda_{v}$) 
coupling in the model. This aspect is clearly demonstrated in our TM1e results, emphasizing the importance of incorporating
both the $\sigma$-cut potential and $\omega$-$\rho$ couplings for better alignment with observational constraints,
particularly for neutron stars with non-nucleonic composition \cite{RIB.19}. 
It is important to constrain the value of $f_{s}$ which is constrained by theoretical studies and 
observational data in a model-independent manner using a Bayesian approach.  


\end{document}